\documentclass[sigconf]{acmart}

\pdfoutput=1

\usepackage{threeparttable}
\usepackage{listings}
\usepackage{enumitem}
\usepackage[frozencache=true, cachedir=minted-cache]{minted}
\usepackage{framed}
\usepackage{color}
\usepackage{bbding}
\usepackage{multirow}
\usepackage{colortbl}
\usepackage{paralist}

\definecolor{mygray}{gray}{.9}

\newenvironment{code}{\captionsetup{type=listing}}{}

\AtBeginDocument{%
  \providecommand\BibTeX{{%
    \normalfont B\kern-0.5em{\scshape i\kern-0.25em b}\kern-0.8em\TeX}}}

\copyrightyear{2022}
\acmYear{2022}
\setcopyright{acmcopyright}\acmConference[ICSE-SEIP '22]{44nd International Conference on Software Engineering: Software Engineering in Practice}{May 21--29, 2022}{Pittsburgh, PA, USA}
\acmBooktitle{44nd International Conference on Software Engineering: Software Engineering in Practice (ICSE-SEIP '22), May 21--29, 2022, Pittsburgh, PA, USA}
\acmPrice{15.00}
\acmDOI{10.1145/3510457.3513055}
\acmISBN{978-1-4503-9226-6/22/05}

\begin{document}

\title{An Industrial Experience Report on Retro-inspection}

\author{Lanxin Yang, He Zhang, Fuli Zhang, Xiaodong Zhang, Guoping Rong}
\affiliation{%
  \institution{State Key Laboratory of Novel Software Technology, Software Institute, Nanjing University}
  \city{Nanjing}
  \state{Jiangsu}
  \country{China}
}
\email{yang931001@outlook.com, hezhang@nju.edu.cn, {mg1932016, mg1932018}@smail.nju.edu.cn, ronggp@nju.edu.cn}

\begin{abstract}
To reinforce the quality of code delivery, especially to improve future coding quality, one global Information and Communication Technology (ICT) enterprise has institutionalized a retrospective style inspection (namely \emph{retro-inspection}), which is similar to Fagan inspection but differs in terms of stage, participants, etc. This paper reports an industrial case study that aims to investigate the experiences and lessons from this software practice. To this end, we collected and analyzed various empirical evidence for data triangulation. The results reflect that retro-inspection distinguishes itself from peer code review by identifying more complicated and underlying defects, providing more indicative and suggestive comments. Many experienced inspectors indicate defects together with their rationale behind and offer suggestions for correction and prevention. As a result, retro-inspection can benefit not only quality assurance (like Fagan inspection), but also internal audit, inter-division communication, and competence promotion. On the other side, we identify several lessons of retro-inspection at this stage, e.g., developers' acceptance and organizers' predicament, for next-step improvement of this practice. To be specific, some recommendations are discussed for retro-inspection, e.g., more adequate preparation and more careful publicity. This study concludes that most of the expected benefits of retro-inspection can be empirically confirmed in this enterprise and its value on the progress to continuous maturity can be recognized organization-wide. The experiences on executing this altered practice in a large enterprise provide reference value on code quality assurance to other software organizations.
\end{abstract}



\maketitle

\section{Introduction}
Quality Assurance (QA) is paramount to software development and therefore has always been highly valued~\cite{laporte2018software}. Code review aims to detect and identify defects in source code, and it has generally become indispensable in ensuring software quality~\cite{fagan1976design, rigby2013convergent}. Besides, code review is beneficial to knowledge transfer, project schedule, and team awareness~\cite{bacchelli2013expectations}, etc. Nowadays, code review has been widely employed in both open source communities and commercial contexts~\cite{bosu2016process, alami2019does, sadowski2018modern}.

Over the past few years, a number of societies and organizations, e.g., IEEE Computer Society~\cite{ieee2008standard}, Google~\cite{sadowski2018modern}, Microsoft~\cite{bosu2015characteristics}, Sony~\cite{shimagaki2016study}, Samsung~\cite{hasan2021using}, and Xerox~\cite{alomar2021refactoring} have reported on their standards, guidelines, and practices with code review. However, despite many benefits and various experiences reported on code review, it remains a challenging practice of QA~\cite{han2020does, fatima2020software, dougan2022towards}. Code review sometimes fails to find defects~\cite{czerwonka2015code}, and worse yet, it slows down workflow~\cite{ebert2019confusion}, results in unfairness~\cite{german2018my}, etc. Compared with other QA practices, e.g., testing, code review largely relies on organizational culture and developers' expertise, experience, and engagement~\cite{baum2016factors, bosu2016process}, which therefore has long puzzled enterprises.

One global Information and Communication Technology (ICT) enterprise has consistently placed the top priority on its product quality as well as engineering capability. As one of the critical QA practices, code review has been adopted in its software development divisions for over three decades. In the early years, code review in this enterprise was mostly conducted in a collective and structured style (known as ``Fagan inspection''~\cite{fagan1976design}), which took place after a source code-related artifact reached the pre-defined exit criteria (e.g., completing specific requirements). Considering such an approach was very time-consuming and cost-intensive, nowadays, the Fagan inspection has been replaced by an asynchronous, lightweight, tool-based, and source code-oriented manner, known as ``peer code review'' or ``modern code review''~\cite{bacchelli2013expectations}. However, the outcomes of peer code review are generally not satisfactory as expected. For instance, some reviewers with limited expertise and experience accordingly have less competence to detect a decent number of defects~\cite{czerwonka2015code}. On the other side, it is likely that there exist cozy relationships among colleagues within the development team and division, which may make peer code review ostensible and biased~\cite{german2018my}. Worse yet, although all the code commits merged into project repositories subject to review and testing, and even work in life, the residual (undetected) defects may consume significant time, effort, and cost to fix and maintain.

Incipiently, a few development teams in this enterprise organized informal postmortems and other types of spontaneous meetings to react to the above challenges, e.g., address critical problems, and more importantly, to seek directions for future improvement. Distinct from Fagan inspection and peer/modern code review, this retrospective style practice turns out to be open-minded by removing the pre-defined entry criteria, i.e. no longer event-driven (like Fagan inspection). In the past few years, the pioneers' experience to some extent confirmed its effectiveness in improving software quality and fostering developer's quality awareness. The positive effect encouraged a number of followers, which attracted management's attention. After surveys on pilot teams and discussion with QA and audit divisions, the practice was officially acknowledged, and then standardized and institutionalized to be an organization-wide practice. In general, its execution is similar to a Fagan inspection but with several alterations. In particular, the inspected projects should have been already reviewed and tested before, and the inspectors must include the technical personnel from the division under inspection. For the sake of distinction and succinctness, the term ``retro-inspection'' refers to this altered Fagan inspection in the rest of this paper unless it is specifically stated otherwise. 

The objective of this study is to investigate this enterprise's experience and lessons on conducting retro-inspection. To this end, we employed a triangulation strategy that aggregates the data from multiple sources for analysis. We first learned retro-inspection from public documents and informal inquiries, then analyzed archive documents from the latest seven retro-inspections (2019 -- 2021). Prior to interviewing the experts who ever participated in retro-inspections, we designed sixteen closed-ended questions in questionnaires to obtain the state of peer code review, so as to prepare appropriate questions for experts. Finally, the question lists consist of fourteen open-ended questions, which are designed to consult experienced inspectors on retro-inspection.

After long-term collections and analyses of various empirical evidence, we understand that through third-party suggestion or/and self-submission, the samples into retro-inspections can be particular projects picked according to the specific interests. A number of domain experts are invited into retro-inspections, in which they have identified quite a few complicated (sometimes even deep-rooted) and underlying code defects, including but not limited to logic faults, security vulnerabilities. Besides, they have provided widespread indicative and suggestive comments, such as the rationale behind the defects, suggestions for correction and prevention. As a result, retro-inspection not only benefits code quality assurance, but also works for internal audit, inter-division communication, and competence promotion. Furthermore, we identify that retro-inspection at this stage is also associated with limitations and risks, especially, synergy among multiple participant roles, which result in lessons such as developers' disagreement and organizers' predicament. In the end, we propose some recommendations for improving retro-inspection, e.g., stricter admission criteria, longer inspection period, and more careful publicity. As an altered QA practice, retro-inspection is still in its stage of continuous improvement.

The main contributions of this industrial case study can be highlighted as follows.
\begin{itemize}
    \item This study reports on an empirical investigation on retro-inspection, as an altered QA practice, in an industrial context through long-term observation with data triangulation.
    \item This study elaborates on retro-inspection as an effective supplement to peer code review, with variations on purposes, stages, samples, participants, and processes.
    \item This study identifies the limitations and potential risks associated with retro-inspection at this stage and discusses recommendations for improvement, which offer reference value for other organizations with serious concerns on software quality.
\end{itemize}

\section{Background and related work}
\subsection{State-of-the-practice of code review}
Since Michael Fagan put forward inspection at IBM in 1976~\cite{fagan1976design}, it has been widely accepted and evolved into multiple forms. The IEEE Computer Society has published its standards for software reviews and audits~\cite{ieee2008standard}, which elaborate on (1) management review, (2) technical review, (3) inspection, (4) walk-through, and (5) audit. Regardless of the types of reviews, they are all required to review source code, i.e. code review is the most fundamental. 

As defined in Google's engineering practices documentation, the code review guidelines\footnote{https://google.github.io/eng-practices/review/} consist of two parts: (1)~\emph{How To Do A Code Review}, and (2)~\emph{The CL Author's Guide}. The former is intended for code reviewers, which elaborates on review standards and emphasis, comment writing, pushback handling, etc. The latter serves developers whose code changes are going through reviews, which elaborates on strategies for preparing submissions and handling review comments. GitLab also offers guidelines\footnote{https://docs.gitlab.com/ee/development/code\_review.html} for performing code review, which elaborate on GitLab-specific context and concerns, multiple roles' responsibilities, approval thresholds, and a number of best practices, etc. Apart from GitLab, there are a variety of tools available to assist in code review, such as Review Board\footnote{https://www.reviewboard.org/} and Gerrit\footnote{https://www.gerritcodereview.com/}. 

When it comes to the enterprises' practices and experiences on code review, an investigation~\cite{bacchelli2013expectations} conducted at Microsoft reveals that in addition to finding defects, code review serves extra benefits such as knowledge transfer, team awareness improvement, and supplementation of alternative solutions to issues. A case study~\cite{sadowski2018modern} conducted at Google shows that all roles highly appreciate code review, seeing it as providing multiple benefits and a venue where one can exchange with others to build, establish, maintain, and evolve norms that ensure readability, integrity, and consistency of project repository. 


\subsection{Limitations and risks of code review}
Code review brings a number of benefits, which have been widely agreed in community~\cite{bacchelli2013expectations, caulo2020knowledge}. Nowadays, code review has generally become a must-have in modern software development. However, despite the benefits of code review, it is subject to various limitations and risks. Fatima et al.~\cite{fatima2020software} identified 28 unique wastes in code reviews, including poor code quality, duplication of work, time spent on code understanding, lack of motivation to share knowledge, etc. Egelman et al.~\cite{egelman2020predicting} pointed out that reviewers' critical examinations may have negative influences, e.g., frustration and stress on colleagues, and ultimately result in their abandonment from projects. German et al.~\cite{german2018my} surveyed developers from OpenStack community, finding that a significant portion of respondents perceive their contributions were unfairly reviewed. They also indicated that there is a lack of consistency and fairness when reviewers prioritize assignments. Han et al.~\cite{han2020does} found that quite a few coding convention violations were ignored by both code authors and reviewers in Eclipse community. Their results indicate that humans are neither effective nor consistent in avoiding convention violations. Bosu et al.~\cite{bosu2015characteristics} found that the proportion of useful comments made by code reviewers increases dramatically in their first year at Microsoft but towards to plateau afterward, which indicates the degeneration of activeness and engagement.

These limitations lead to the risks that turn code review to be a waste of time. With the emphasis on code quality, the global ICT enterprise studied in this paper has conducted a retrospective inter-division style code review, namely \emph{retro-inspection}. The experiences and lessons learned from retro-inspections are reported in this paper.

\section{Preliminary: retro-inspection}
\label{SEC:PRE}
This section describes retro-inspection protocols, including the purposes, stages, samples, participants, and processes \& activities.

\begin{table*}[htbp]
\vspace{-2.0ex}
\centering
\footnotesize 
\caption{Comparison of common types of software inspections and audit} 
    \label{TAB:reviewComparsion}
    \vspace{-3.0ex}
\begin{threeparttable}
\begin{tabular}{p{15mm}p{30mm}p{38mm}p{40mm}p{38mm}}
\midrule[1pt]
\hline
\rowcolor[gray]{0.9} & Software audit & Classical inspection (Fagan style) & Modern inspection (modern code review) & \textbf{Retro-inspection (enterprise's)} \\
\hline
Objective & internal audit & project scheduling, quality assurance & quality assurance & internal audit, quality assurance \\
Target & software processes\tnote{1}, outputs & software outputs\tnote{2} & source code & source code \\ 
Venue & outside & inside & inside & outside \\
Stage & any stages & any stages & pre-merge & post-test \\
Scope & partial & partial & global & partial \\
Frequency & yearly & weekly/monthly & daily & quarterly \\
Communication & meeting & meeting & comment & comment, meeting \\ 
Role & auditor, audited organization & inspector, inspected organization & reviewer, reviewed organization & inspector, inspected organization \\
\hline
\midrule[1pt]
\end{tabular}
 \begin{tablenotes}
        \footnotesize
        \item[1] Examples of software process-related documentations include, but are not limited to (1) standards, (2) regulations, (3) guidelines, (4) plans, and (5) procedures.
        \item[2] Examples of software outputs/products include, but are not limited to (1) requirements specifications, (2) design descriptions, (3) test documentations, and (4) source code.
      \end{tablenotes}
    \end{threeparttable}
    \vspace{-3.0ex}
\end{table*}

\textbf{Purposes:}
Retro-inspection in this enterprise is not limited to tackling current quality issues with source code, instead it has great significance in improving future quality and fostering developers' awareness of quality concerns.

\begin{figure}[!htbp]
\vspace{-1.0ex}
  \centering
  \includegraphics[width=0.95\linewidth]{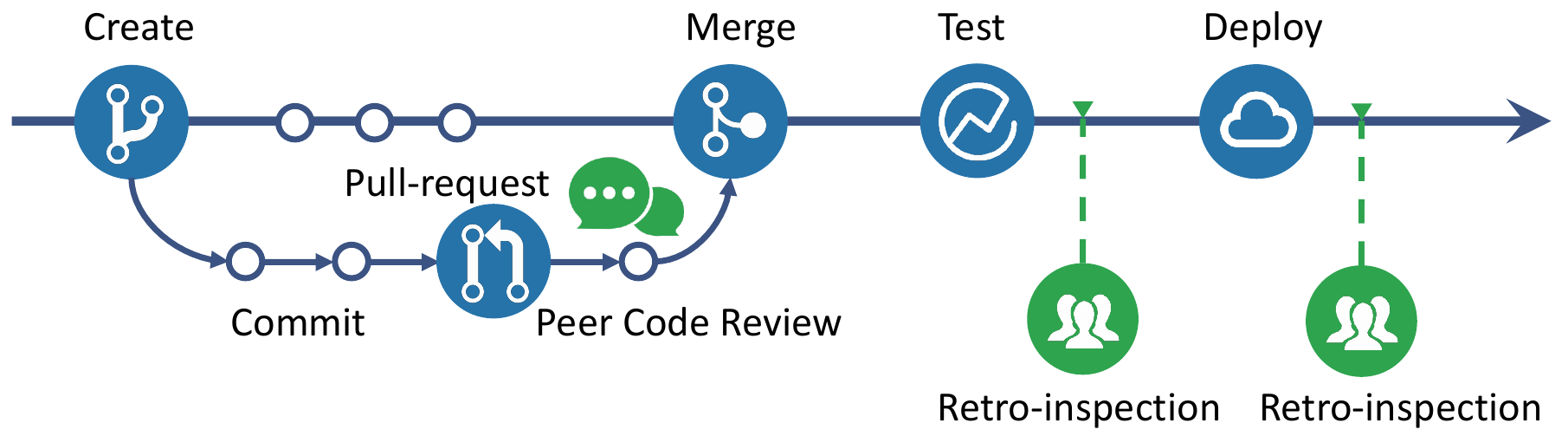}
  \vspace{-1.0ex}
  \caption{Retro-inspection in code pipeline}
  \label{FIG:inspectionStage}
  \vspace{-1.0ex}
\end{figure}

\textbf{Stages:}
Figure~\ref{FIG:inspectionStage} shows an overview of peer code review and retro-inspection in the enterprise's code pipeline. Peer code review is mandatory within divisions prior to merging code commits into project repositories, whereas retro-inspection is optional and should be conducted by external divisions after testing has been completed. 

\textbf{Samples:}
The samples of retro-inspection consist of a wide range of projects, including (but are not limited to) (1) problem-prone projects, (2) newly-deployed projects, (3) projects that support core business functions, (4) outsourced projects, and (5) projects never inspected before. Besides, (6) well-performed projects are also allowed to be picked according to specific interests. Managements make the choice of inspection samples based on (1) third-party recommendations (e.g., security and QA divisions), and/or (2) self-submission by inspected divisions. 

\textbf{Participants:}
Multiple roles participate in retro-inspections, and each role is responsible for specific tasks. The executives are the leaders and decision-makers. The organizers are responsible for inspection protocols, personnel coordination, process monitoring, and reporting. The chief inspectors who are both coding and inspection experts are in charge of personnel coordination and process monitoring. Additionally, 2 -- 4 coding experts are assigned to one group to conduct inspection. Note that each inspector should possess rich experience in peer code review but does not necessarily require prior experience on retro-inspection. Finally, the inspected divisions' related roles, such as code authors, are responsible for defect correction and quality improvement. 

\begin{figure}[!htbp]
  \vspace{-2.0ex}
  \centering
  \includegraphics[width=0.82\linewidth]{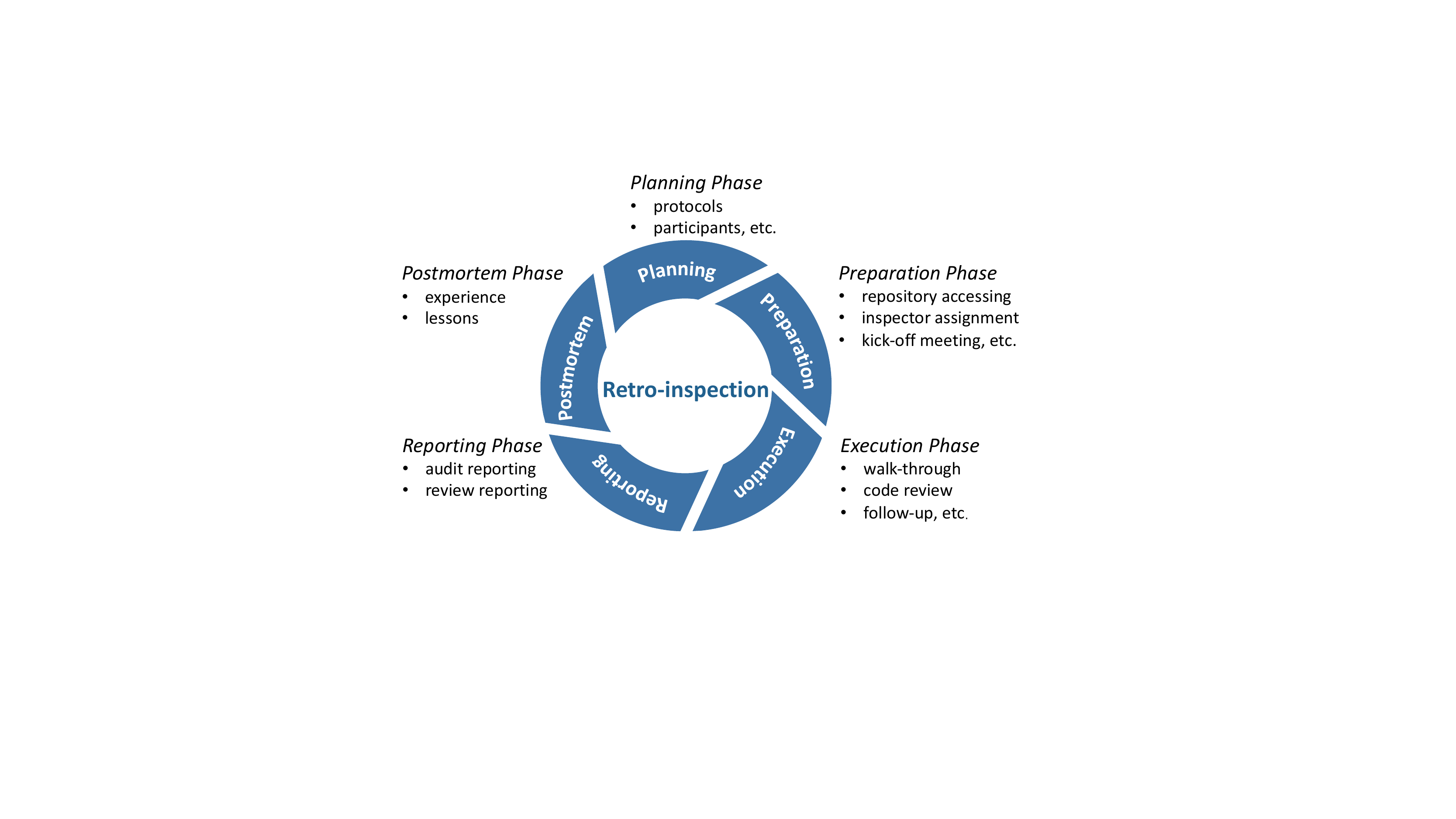}
  \vspace{-1.0ex}
  \caption{Processes \& activities of retro-inspection}
  \label{FIG:inspectionProcess}
  \vspace{-2.0ex}
\end{figure}

\textbf{Processes \& activities:}
Figure~\ref{FIG:inspectionProcess} shows an overview of the processes \& activities of retro-inspection. In the planning phase, executives determine objectives and the inspected divisions, and organizers specify inspection protocols, and participants. In the preparation phase, organizers access inspected divisions and samples, assign inspectors, and host kick-off meetings. In the execution phase, organizers and the inspected divisions take charge of walk-throughs; inspectors collaborate to examine samples in either an asynchronous or synchronous manner, and communicate with code authors if needed; code authors continuously improve and monitor code changes until all the identified defects are fixed and confirmed with inspectors. There are few differences between retro-inspection and peer code review in the execution phase. In the reporting phase, organizers conclude results and findings, and report to managements, chief inspectors, and the inspected divisions. Finally, all participant roles share experiences and lessons in the postmortem phase. For instance, \emph{whether or not the code defects found in retro-inspections can be detected and identified by static analysis tools?} (If yes, peer code review and testing should be improved; if not, static analysis tools should be strengthened.)

As a variation of Fagan inspection, retro-inspection in this enterprise is conducted by personnel out of the division under inspection, and the samples should have been reviewed and tested in advance. Retro-inspection is similar to software audit~\cite{ieee2008standard} but to source code, it has no pre-defined criteria for other types of software outputs/products as well as software processes. Also, retro-inspection shares a very close relationship with peer code review, but the differences are also observable. More comprehensive comparisons of common types of inspections and audit are presented in Table~\ref{TAB:reviewComparsion}.

\section{Research method}
\begin{figure*}[!htbp]
  \centering
  \includegraphics[width=0.90\linewidth]{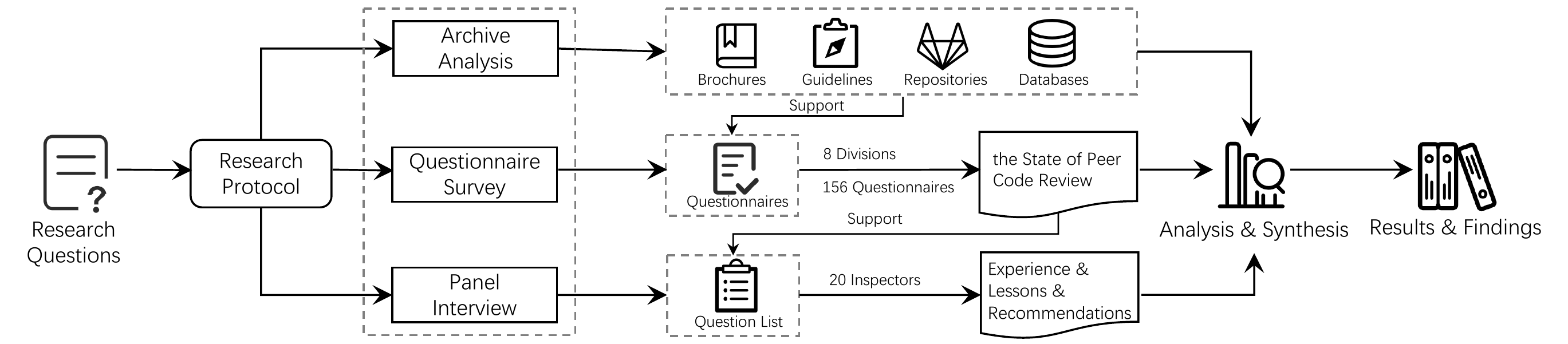}
  \vspace{-1.0ex}
  \caption{An overview of research method and process}
  \label{FIG:researchMethod}
  \vspace{-3.0ex}
\end{figure*}

Following the research design guidelines~\cite{creswell2017research}, we conducted a case study aiming to report on retro-inspection in an industrial context. Figure~\ref{FIG:researchMethod} shows an overview of the research method and process. 

\subsection{Research questions}
The four research questions (RQs) that were proposed to guide this industrial case study are as follows.
\begin{itemize}
  \item \textbf{RQ1:} What are the benefits of retro-inspections?
  \item \textbf{RQ2:} What comments are given in retro-inspections?
  \item \textbf{RQ3:} What defects are identified in retro-inspections?
  \item \textbf{RQ4:} What lessons are learned from retro-inspections?
\end{itemize}

As the enterprise-specific software practice, retro-inspections require well-defined protocols (cf. Section~\ref{SEC:PRE}). We propose RQ1 to explore the benefits of retro-inspections, from both management's expectations and developer's perceptions. RQ2 and RQ3 aim to investigate the outputs (in terms of comments and defects) that contribute value for retro-inspections. 
Finally, RQ4 is promoted to aggregate lessons from retro-inspections, so as to support its future improvements.


\subsection{Data collection}
Retro-inspection in this enterprise is closely related to peer code review and other software practices, such as development and testing. Understanding retro-inspection requires long-period participation and observation. Since July 2020, one author had worked as a full-time intern in this enterprise, and the other two authors visited it periodically for observation.

Due to access restrictions, unfortunately, we can only access limited repositories, databases, and participants involved in retro-inspections. Data triangulation is a research strategy that develops a comprehensive understanding of phenomena by using multiple data sources~\cite{thurmond2001point}. By using triangulation, we can make the most of each single data source's advantages meanwhile overcoming shortcomings to some extent. 

\subsubsection{Archive analysis}
The preliminary understanding of retro-inspection comes from reading (1) brochures, (2) guidelines, followed by retrieving information from (3) repositories and databases. The first two help to understand inspection protocols, such as participants, processes \& activities. The last helps to analyze inputs and outputs of retro-inspection, such as samples' fields and inspectors' outputs. 

Review comments convey rich information~\cite{bosu2015characteristics} for archive analysis. We manually analyzed the latest seven retro-inspections' comments (2019 -- 2021, in 3292 total). From them, we identified and summarized comment styles and code defects. (cf. Section~\ref{SEC:ANA} for analysis methods and processes.)

\subsubsection{Questionnaire survey}
Retro-inspection is not only an inspection of source code but also peer code review since all the samples should have already been reviewed as a regular basis. To make up for the limited understanding of peer code review, and to develop appropriate questions for panel interviews, we conducted an online survey of developers who have participated in peer code reviews. The questionnaire consists of sixteen closed-ended questions. Among them, the first three are general questions that help to distinguish interviewees' roles, understand their perception of review benefits, and communication tools. Then two questions (i.e. feedback period and comment styles) are designed for code authors only, the rest of the questions (e.g., review workload, review habits, time spent, etc.) for code reviewers only. The complete questionnaire is available in the online appendix\footnote{http://softeng.nju.edu.cn/tech-reports/TR-22-002-CodeReview-EN.pdf}.

We identified the most productive code authors (in terms of their commit records) and reviewers (in terms of their review outputs) in the enterprise, and choose their intersection as the target population for the questionnaire survey. In the fourth quarter of 2020, 200 questionnaires were distributed to them using an internal system. The requested response time was given for three days. In the end, 156 valid questionnaires were returned from eight divisions with an overall response rate of 78\%. 


\subsubsection{Panel interview}
To achieve deep insights into retro-inspection, we conducted panel interviews~\cite{dixon2002panel} with experts who have ever participated in it before. We designed fourteen basic open-ended questions. Among them, the first six questions help to understand experts' demographics and their experiences (e.g., participation times, roles, workload assignment). Subsequent three questions (e.g., inspection emphasis and comment styles) are designed for accessing experts' habits. The remainders are proposed to aggregate suggestions for improvement (e.g., differences from peer code review, benefits, shortcomings, and suggestions). We adjusted questions' order and supplemented questions according to interviewees' responses. The basic questions for panel interviews are available in the online appendix, accessible with the questionnaire. 

All the interviews were recorded with the interviewees' permission. Three authors took part in panel interviews. Two of them asked questions in turn, the other was in charge of recording. In total, we interviewed twenty inspectors. Twelve of them are security experts, the others are business and functional experts. Note that each role can freely comment on any type of code defect. 


\begin{figure}[!htbp]
  \centering
  \vspace{-1.0ex}
  \includegraphics[width=0.9\linewidth]{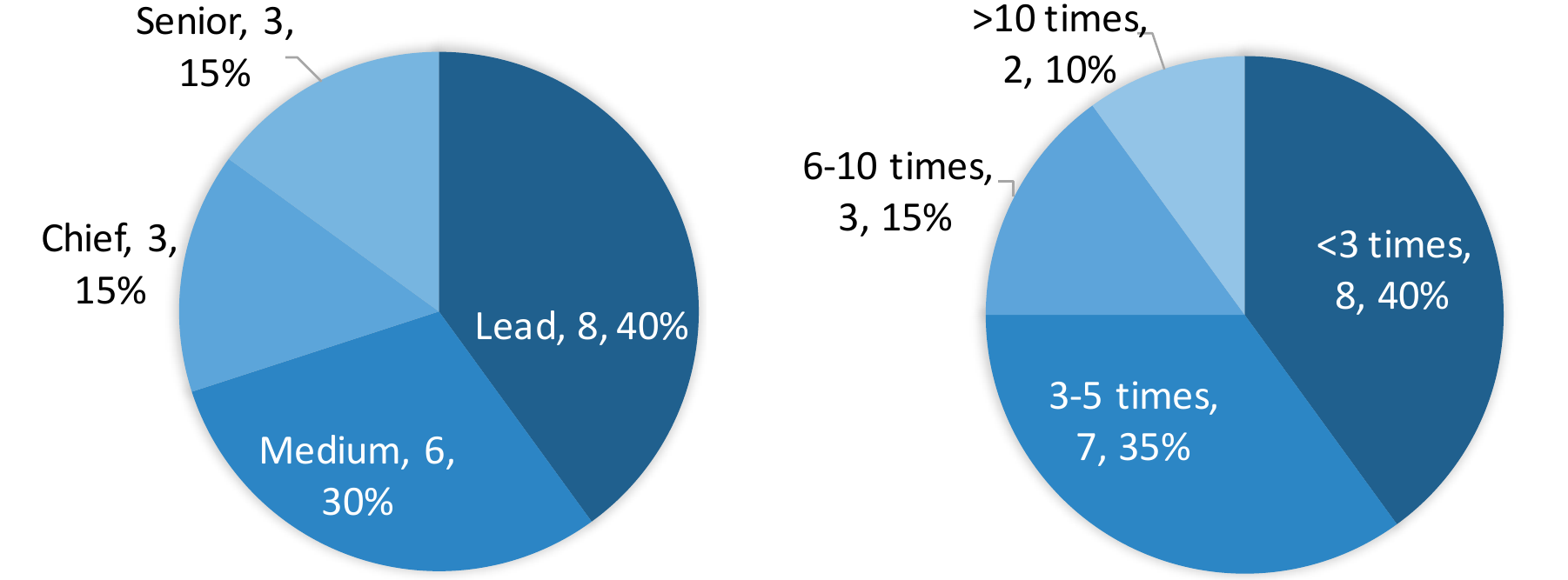}
  \caption{Demographics of the interviewed inspectors (Left: position; Right: participation times)}
  \vspace{-2.0ex}
  \label{FIG:inspectors}
\end{figure}

Figure~\ref{FIG:inspectors} shows the demographics of the interviewed inspectors. The inspector's position ranges from \emph{Medium-level Engineer} ($ME$) to \emph{Chief Engineer} ($CE$), indicating their working experience in the enterprise and their qualification in a specific area, in particular on peer code review. As shown in the figure, over half (12/20) of the inspectors has participated in retro-inspections for at least 3 times. 
In the rest of this paper, each participating inspector is denoted by his/her position and inspection experience. For instances, $LE_{[7]}$ denotes a \emph{Lead Engineer} with 7 times experience on retro-inspection. 
Besides, in our study with this enterprise, we frequently communicated with two management staff who took the role of organizers (indicated by $O_1$ and $O_2$) in retro-inspections. 

\subsection{Data analysis}
\label{SEC:ANA}
Both quantitative and qualitative data collected in this study went through the following methods for analysis to generate findings.

\paragraph{Descriptive analysis} a.k.a. descriptive statistics, is a quantitative analysis method that uses statistical methods to describe and summarize data~\cite{de2019evolution}. The archive documents consist of a large amount of data from both peer code review and retro-inspection, and our questionnaires consist of several closed-ended questions. By analyzing the up-to-date state of review processes and results with descriptive analysis methods, we can develop a preliminary understanding of peer code review, and further design specific questions for understanding retro-inspection.

\paragraph{Thematic analysis} is used as a qualitative analysis method to identify common themes within data~\cite{braun2006using}. It mainly helps to understand major comment styles (RQ2) and code defects (RQ3). Coding is the process of labeling and organizing qualitative data to identify and distinguish themes and the relationship between them~\cite{saldana2021coding}, and was therefore utilized in thematic analysis. Although the inspection comments have been labeled with defect types and severity levels, they are incomplete and follow inconsistent taxonomies (mainly for defect types). Therefore, we re-labeled comments from the latest seven retro-inspections. We first learned the enterprise's specifications, inquired developers to understand common defects. Next we checked the existing labels from a small amount (300) of comments to identify major (and recurring) labels. Then two authors independently assigned labels to each comment. Finally three authors collaborated to review and reassign labels until reach a consensus. For example, given the inspection comment -- ``\emph{Please move this to /secureconfigs/secureconfigs.properties file with permission as read-only for this user and 0 for other groups}'', its defect type is labeled as ``\texttt{Sec}'' (security vulnerability), and its severity level as ``\texttt{Maj}'' (major), and its comment style as ``\texttt{Sug}'' (serving suggestions). 

\paragraph{Narrative inquiry or narrative analysis} is used to qualitatively inquire and analyze the stories people engaged~\cite{connelly1990stories}. This method helps to understand how interviewees represent themselves and their experiences in retro-inspections. In order to acquire inspector's understanding of retro-inspection, the interviewees were encouraged to share their most unforgettable experiences, views of benefits, shortcomings as well as suggestions for inspection process improvement. All interviews were recorded and then transcribed. Three authors collectively performed coding for each interview in case of misunderstanding.

\section{Findings and lessons}
\label{SEC:RES}
This section elaborates on the results and findings that contribute to answering the research questions. 

\subsection{Benefits of retro-inspections (RQ1)}
\label{SEC:BEN}
Four main benefits of retro-inspections are identified from inquires.

\paragraph{\textbf{Benefit 1: Quality assurance.}}
As its primary objective, code review is expected to ensure code quality~\cite{bacchelli2013expectations}. However, peer code review is often conducted in rush due to the schedule pressure and the reviewer's limited expertise. Although all the code commits merged into project repositories have been reviewed, tested, and even work in production, there may exist missing defects. Worse yet, some of them are at great risk. Retro-inspection aims to further detect and identify them, as well as seek opportunities for prevention or alteration, which is highlighted in its motto -- ``\emph{Look Back to Look Forward: The Way of Success}''. The inspection organizer ($O_1$) introduced that ``\emph{Distinct from peer code review that is under heavy business pressure, and is limited to pre-defined criteria/checklists, retro-inspection has few limitations}''. Moreover, various software professionals, such as architects, testing specialists, and security experts have collaborated in retro-inspections, bringing plenty of their own experiences, which makes the evident improvement on code quality. (cf. Section~\ref{SEC:RQ3} for quantitative analysis.)

\paragraph{\textbf{Benefit 2: Internal audit.}}
Internal audit is an independent inspection, supervision, and evaluation activity that provides an unbiased view of the audited population~\cite{behrend2019evolution}. It helps to identify and prevent risks, make decisions, strengthen management, etc. Retro-inspection is an unannounced inter-divisional examination that supports the enterprise's internal audit. Code quality, as one of the major measures to evaluate development quality, therefore, becomes the emphasis of retro-inspection. Besides, since all projects to be examined in retro-inspections should have already been reviewed and tested before, retro-inspection also contributes to evaluating the quality of the other QA practices (e.g., review and testing). 

\paragraph{\textbf{Benefit 3: Inter-division communication.}}
Quite a few new divisions have been set up in this enterprise in recent years. Most of them have their own business and the responsibility gaps between divisions gradually become evident. However, they often request or provide services on/to others. Inter-division communication, especially on understanding business backgrounds and service functions, has become necessary. 
Retro-inspection serves as a great opportunity for inter-division communication and collaboration, in which inspectors can not only discuss the projects under inspection, but also the business and services they have worked with. An executive indicated that he considers experts whom he worked with in retro-inspections as candidates for new projects.

\paragraph{\textbf{Benefit 4: Competence promotion.}}
By conducting long-term and large-scale retro-inspections, a large number of cases, either poorly- or well-implemented, have been identified, aggregated, and archived. From them, QA divisions extracted commonly occurring defects and then schedule periodic capability examinations, training courses, and expert consulting for developers accordingly. Also, the inspectors who participated in retro-inspections are invited to be tutors, examiners ($O_1$). Moreover, retro-inspection can also benefit the inspectors themselves. Inspectors are mostly domain experts who possess both expertise as well as experience, and retro-inspections are conducted in a group- or (at least) pair-review manner, which makes peer communication dominant in retro-inspections. Thereafter, retro-inspections provide a unique opportunity to cultivate professionals. Four inspectors ($LE_{[7]}$, $ME_{[10+]}$, $ME_{[10+]}$, $ME_{[5]}$) indicated that they volunteered to participate in retro-inspections to advance their expertise. Other inspectors ($CE_{[1]}$, $LE_{[1]}$, $LE_{[3]}$) indicated that everyone within their division is required to alternately take part in retro-inspections.

\vspace{-1.0ex}
\begin{framed}
\vspace{-1.0ex}
\textbf{Finding:} In addition to its major function -- code quality assurance (especially for future quality), retro-inspection can contribute to side benefits such as internal audit, inter-division communication, and additionally, plays a role in cultivating developers and reviewers. 
\vspace{-1.0ex}
\end{framed}
\vspace{-1.0ex}

\subsection{Comments in retro-inspections (RQ2)}
As several software experts who get used to peer code review are invited to a retro-inspection, we are interested in their specific considerations in retro-inspections, which may help explain and transfer the success of retro-inspection. This subsection reports the styles of inspection comments from the content analysis.

\paragraph{\textbf{Comment 1: Clarifying defect.}}
Detecting and identifying defects is the most primary task of code review. All the interviewed inspectors indicated that they clarified defects in retro-inspection comments. For anything unclear, they would contact code authors or initiate a discussion within the inspection group. A few interviewees (5.13\%) who participated in questionnaire surveys indicated that they sometimes raised queries in peer code reviews. Hence, comments in retro-inspections serve for easy understanding, while comments in peer code reviews can work for team communication.

\paragraph{\textbf{Comment 2: Elaborating rationale.}}
Brief assessments are common in peer code reviews, such as ``\emph{incorrect logic}'', ``\emph{unnecessary}''. Beyond indicating defects, code authors, especially the newcomers, also need to know the rationale behind for correction and further improvement. Note that the `rationale' in this context implies ``\emph{why does it fail to function}'', rather ``\emph{where make it break}'', or ``\emph{why offer such suggestions}''. In retro-inspections, inspectors elaborate the rationale for avoiding misunderstanding and disagreements. 
All the interviewed inspectors confirmed that they were willing to elaborate the rationale in retro-inspections. On the contrary, only 43.59\% of the interviewees offer the rationale in peer code reviews. The interviewees indicated that in the context of peer code reviews, where they share a similar background with code authors, comments contain only concise explanation for minor defects, while major defects can be further discussed via voice calls and face-to-face meetings.

\paragraph{\textbf{Comment 3: Offering suggestion.}}
Apart from indicating defects, the ultimate objectives of code review are correction, prevention, and improvement. Listing~\ref{CODE:badCode} shows a case of poor-quality Java code identified in retro-inspections. The defect description indicates that lines 6 -- 12 misuse~\texttt{synchronized} when printing log messages. In addition to indicating the defect, the improvement is also suggested --``\emph{No need of synchronization if no DATA-RACE; use DOUBLE-CHECK if lock necessary}''. 80.00\% of the interviewed inspectors confirmed that they had served suggestions in comments. When it comes to peer code reviews, only 28.85\% of interviewees offered suggestions in comments. Instead, 83.33\% of interviewees used to sharing suggestions by casual communications. 

\begin{code}
\vspace{-1.0ex}
\captionof{listing}{A case of low-quality code}
\vspace{-1.0ex}
\label{CODE:badCode}
\begin{minted}[frame=lines, fontsize=\scriptsize, bgcolor=white, numbersep=-1pt, highlightlines={6,7,8,9,10,11,12}, linenos=true, showtabs=false, tabsize=4, breaklines]{java}
public static GsLog getDebugLog(String moduleName) {
    if (null == moduleName || moduleName.trim().length() == 0) {
        return null;
    }
    GsLog gsLog = null;
    synchronized (LOCK_LOG) {
        gsLog = (GsLog) logMap.get(moduleName);
        if (null == gsLog) {
            gsLog = new GsLogImpl(moduleName);
            logMap.put(moduleName, gsLog);
        }
    }
    return gsLog;
}
\end{minted}
\vspace{-4.0ex}
\end{code}

\paragraph{\textbf{Comment 4: Appreciating good-job.}}
There is another type of comment that works for developers' cultivation. When coming across excellent designs or implementations in the inspected samples, inspectors express appreciations. Taking Listing~\ref{CODE:goodCase} as a case, the strengths of this Java code snippet include (1) accelerating access by locally caching the users and groups' relationship rather than remotely accessing IAM service through HTTP RESTful interface every time; (2) improving flexibility by setting cache expiration time to reload updated data; (3) avoiding out of memory by setting the maximum number of caches. Six inspectors ($LE_{[7]}$, $LE_{[1]}$, $LE_{[1]}$, $ME_{[5]}$, $ME_{[4]}$, $ME_{[3]}$) indicated that they had shared the masterpieces within their divisions. Having understood its benefits, in recent years, the managements motivate each inspector to report at least one poor case and one good case in a retro-inspection. By contrast, good exemplars found in peer code reviews are more commonly acknowledged in casual communications.

\begin{code}
\captionof{listing}{A case of high-quality code}
\vspace{-1.0ex}
\label{CODE:goodCase}
\begin{minted}[frame=lines, fontsize=\scriptsize, bgcolor=white, numbersep=-1pt, linenos=true, showtabs=false, tabsize=4, breaklines]{java}
private LoadingCache<String, Set<String>> userGroupSet = CacheBuilder.newBuilder()
    .maximumSize(MAX_CACHE_SIZE)
    .expireAfterWrite(DURATION_10, TimeUnit.MINUTES)
    .build(new CacheLoader<String, Set<String>>() {
        @Override
        public Set<String> load(String userId) throws Exception {
            HashMap<String, HashSet<String>> userGroupsMap = IAMScopeQuery.getUserGroupsFromIam(userId, config);
            return userGroupsMap.get(userId);
        }
    });
\end{minted}
\vspace{-4.0ex}
\end{code}

\vspace{-1.0ex}
\begin{framed}
\vspace{-1.0ex}
\textbf{Finding:} Comments in retro-inspections clarify code defects, elaborate rationale behind, and offer suggestions to assist correction and improvement; as well as appreciate well-done work. Despite some of them also appear in peer code reviews, they become more common outcomes of the comments from retro-inspections. 
\vspace{-1.0ex}
\end{framed}
\vspace{-1.0ex}

\subsection{Defects found in retro-inspections (RQ3)}
\label{SEC:RQ3}
According to the inspection protocols, all the samples for inspection should have been reviewed and tested before. However, having checked the inspection reports, we found quite a few defects escaping from prior QA practices. This subsection summarizes these defect types and their severity distribution to further investigate the outputs from retro-inspections.

\subsubsection{Defect types}
Four main types of defects are as follows.

\paragraph{\textbf{Defect 1: Logic faults (Log).}}
The experts who participated in retro-inspections with high levels of coding skills may have a greater chance of detecting and identifying those hard-to-find logic faults remaining in the inspected samples. In the past seven retro-inspections, inspectors had found a total of 193 logic faults, mainly including  business logic faults and functional logic faults. Most of them resulted from poor designs, incorrect or low-efficiency implementations, which should be corrected right away to prevent systems from malfunctioning. 

In the panel interviews, one senior engineer ($SE_{[2]}$) mentioned that ``\emph{to identify defects in code reviews is all well motivated, however, some defects should be examined by machines rather than figured out by human brains. The most critical task in code reviews, especially in retro-inspections, is to identify logic faults}''. However, another lead engineer ($LE_{[3]}$) argued that ``\emph{it is hard to detect and identify business logic faults in retro-inspections since we have little understanding of the project background}''. Although identifying logic faults in retro-inspections is strongly recommended, the associated challenge may result in disagreements (cf. Section~\ref{SEC:RQ4}).

\paragraph{\textbf{Defect 2: Security vulnerabilities (Sec).}}
Security inspection is one of the primary tasks of retro-inspection since it is challenging but critical to software quality. One organizer ($O_1$) indicated that ``\emph{Both developers and reviewers have weak abilities in identifying and preventing security vulnerabilities and therefore largely rely on scanning tools. However, the `False Negatives' and `False Positives' are commonly occurrences}''. Therefore, besides the business and functional experts, each inspection group should include at least one security expert. In the past seven retro-inspections, inspectors had identified a total of 386 security vulnerabilities which are mainly concerned with (1) improper input validation, (2) numeric errors, (3) improper implementation of API contract, (4) permissions, privileges, and access controls, (5) information exposure, etc. 

\paragraph{\textbf{Defect 3: Coding standard violations (Sta).}}
Compliance with coding standards/specifications is essential in any large enterprise, which is beneficial to both software maintainability and team communication. In the past seven retro-inspections, inspectors identified a total of 958 defects related to coding standards, mainly including (1) design specifications, (2) coding styles, (3) variables and types, (4) exception handling, (5) log printing, (6) multi-thread concurrent. Beyond division-specific coding standards, it is even more important to conform higher level enterprise-wide standards. 
Unfortunately, the majority of these violations are marked as `OPTIONAL' in peer code reviews, which implies a low priority with less chance to be finally fixed. 

\paragraph{\textbf{Defect 4: Coding experience violations (Exp).}}
There are a number of commonly accepted practices for high-quality, high-efficiency coding, which share significant amount of commonalities with the coding standards/specifications but largely rely on developers' skill, preference, and experience. The violations to these experiences may result in (1) useless code, (2) duplicate code, (3) long methods/classes, (4) parameter judgment omissions, (5) pointer judgment omissions, etc. Many defects are code smells or idiomatic-against-usages that cover a wide range of cases. Developers, especially newcomers, may fail to address these defects. Therefore, coding experience violations become the most common defects identified in retro-inspections. In the past seven retro-inspections, inspectors identified a total of 1755 defects related to coding experience, up to 53.31\% in all the detected defects. In the near future, these common defects will be covered in the enterprise's coding standards/specifications.

\subsubsection{Defect severity}
In addition to specifying defect types, inspectors also need to indicate their severity in retro-inspections.
\begin{itemize}
    \item \textbf{Major.} The code implementations fail to fit requirements or use cases. Worse yet, they may result in function failures, security vulnerabilities and etc.
    \item \textbf{Moderate.} They are similar to major defects but do not have a significant impact on the functionality and security, or there are better alternatives.
    \item \textbf{Minor.} The issues cause inconvenience of understanding, use, etc., but do not affect functionality and security.
\end{itemize}

Table~\ref{TAB:defectAndSecurity} summarizes the defects identified in the past seven retro-inspections. From a total of 2493.7K Lines Of Code (KLOC), inspectors identified 527 major defects, 1683 moderate, and 1082 minor ones, which may reflect their skills. The dominance of major/moderate defects may imply the low quality of the inspected samples as well as the developers' low quality awareness. It was concluded in the end of panel interviews that ``\emph{It is important to consider why they (defects) remain; how to correct them; and most importantly, how to avoid them. They have significant value of retro-inspection}''.

\begin{table}[htp]
\vspace{-1.0ex}
\centering
\scriptsize
\caption{Distribution of the identified defects (2019 -- 2021)}
\vspace{-2.0ex}
    \label{TAB:defectAndSecurity}
    \small
    \setlength{\tabcolsep}{2.0mm}
\begin{threeparttable}
\begin{tabular}{l|r|rrrr|rrr}
\hline
 & \multirow{2}{*}{kLOC} &
\multicolumn{4}{c|}{Defect} &
\multicolumn{3}{c}{Severity}\\
\cline{3-9} 
& & Log & Sec & Sta & Exp & Maj & Mod & Min\\
\hline
2019 Q1 & 95.02 & 5 & 24 & 81 & 58 & 6 & 77 & 85\\
\hline
2019 Q2 & 221.36 & 15 & 41 & 177 & 273 & 66 & 226 & 214\\
\hline
2019 Q3 & 550 & 67 & 30 & 142 & 437 & 141 & 384 & 151\\
\hline
2019 Q4 & 353 & 14 & 9 & 38 & 103 & 6 & 72 & 86\\
\hline
2020 H1\tnote{*} & 425.90 & 31 & 70 & 145 & 211 & 78 & 232 & 147\\
\hline
2020 H2 & 368 & 48 & 100 & 125 & 234 & 64 & 271 & 172\\
\hline
2021 H1 & 480.42 & 13 & 112 & 250 & 439 & 166 & 421 & 227\\
\hline
Total & 2493.70 & 193 & 386 & 958 & 1755 & 527 & 1683 & 1082\\ 
\hline
\end{tabular}
 \begin{tablenotes}
        \footnotesize
        \item[*] Due to the COVID-19 pandemic, the quarterly retro-inspection has been rearranged to be twice a year.
      \end{tablenotes}
    \end{threeparttable}
    \vspace{-3.0ex}
\end{table}

\vspace{-1.0ex}
\begin{framed}
\vspace{-1.0ex}
\textbf{Finding:} Multiple software experts collectively examine the sampled project in retro-inspections, identifying a wide range of code defects, such as (1) logic faults, (2) security vulnerabilities, as well as violations of (3) coding standards and (4) coding experience. Quite a few of them are serious and should be highly concerned, which confirms the unique value of retro-inspection. 
\vspace{-1.0ex}
\end{framed}
\vspace{-1.0ex}

\subsection{Lessons from retro-inspections (RQ4)}
\label{SEC:RQ4}
Although the retro-inspection protocols are well defined, their execution remains a challenge to the enterprise. This subsection focuses on the lessons learned from retro-inspections.

\paragraph{\textbf{Lesson 1: Developer's acceptance.}}
In most cases, inspectors specify defects and reach an agreement with code authors. For major and moderate defects, however, it is not always an easy job. Whenever code authors do not agree on the defects or their severity, they may undergo a long period and even fierce debate, especially if the defects are closely related to the logic implementations. In the absence of a deep understanding of business or functional logic, inspector's comments may not be effective. On the other side, some code is hard to change immediately, and therefore the code authors may have different view on their severity. Moreover, the inspected organization bear tremendous pressures of the evaluation, so as to incline to struggle against inspector's comments. As a result, inspectors turn to be moderate ($LE_{[3]}$, $LE_{[3]}$, $ME_{[4]}$), i.e. some of them pay less attention to serious quality attributes, but emphasize style-oriented flaws, which makes retro-inspection diverging from its origin.    

\paragraph{\textbf{Lesson 2: Inspector's engagement.}}
In a retro-inspection, the participating inspectors' engagements and outputs may significantly differ from each other. Take the latest retro-inspection (2021-H1) as an example, the most productive inspector identified as many as 112 defects (14 \emph{major} + 78 \emph{moderate} + 20 \emph{minor}). On the contrary, the least productive inspector failed to identify any defect because of the suddenly increased workload in his division. While each retro-inspection lasts for approximately two weeks, the outputs are not satisfactory. Several inspectors ($CE_{[1]}$, $LE_{[1]}$, $LE_{[3]}$) explained that they were assigned to the retro-inspection. Yet, their routine workload remain a substantial amount, leading to very few spare moments on inspection. Consequently, it is difficult to guarantee both the quality and the productivity of retro-inspection.

Defect Index (DI), which originally measures software quality~\cite{menzies2008automated} (as shown in Formula~\ref{EQ:DI}), is used in this enterprise to measure inspectors' outputs (and their engagement to some extent) in retro-inspections. 

\vspace{-1.0ex}
\begin{equation}
\vspace{-1.0ex}
\label{EQ:DI}
    \texttt{$DI$} = 3 \cdot \texttt{$N_{maj}$} + 1 \cdot \texttt{$N_{mod}$} + 0.1 \cdot \texttt{$N_{min}$}
\vspace{-1.0ex}
\end{equation}
\vspace{-1.0ex}

\begin{figure}[!htbp]
\vspace{-2.0ex}
  \centering
  \includegraphics[width=0.8\linewidth]{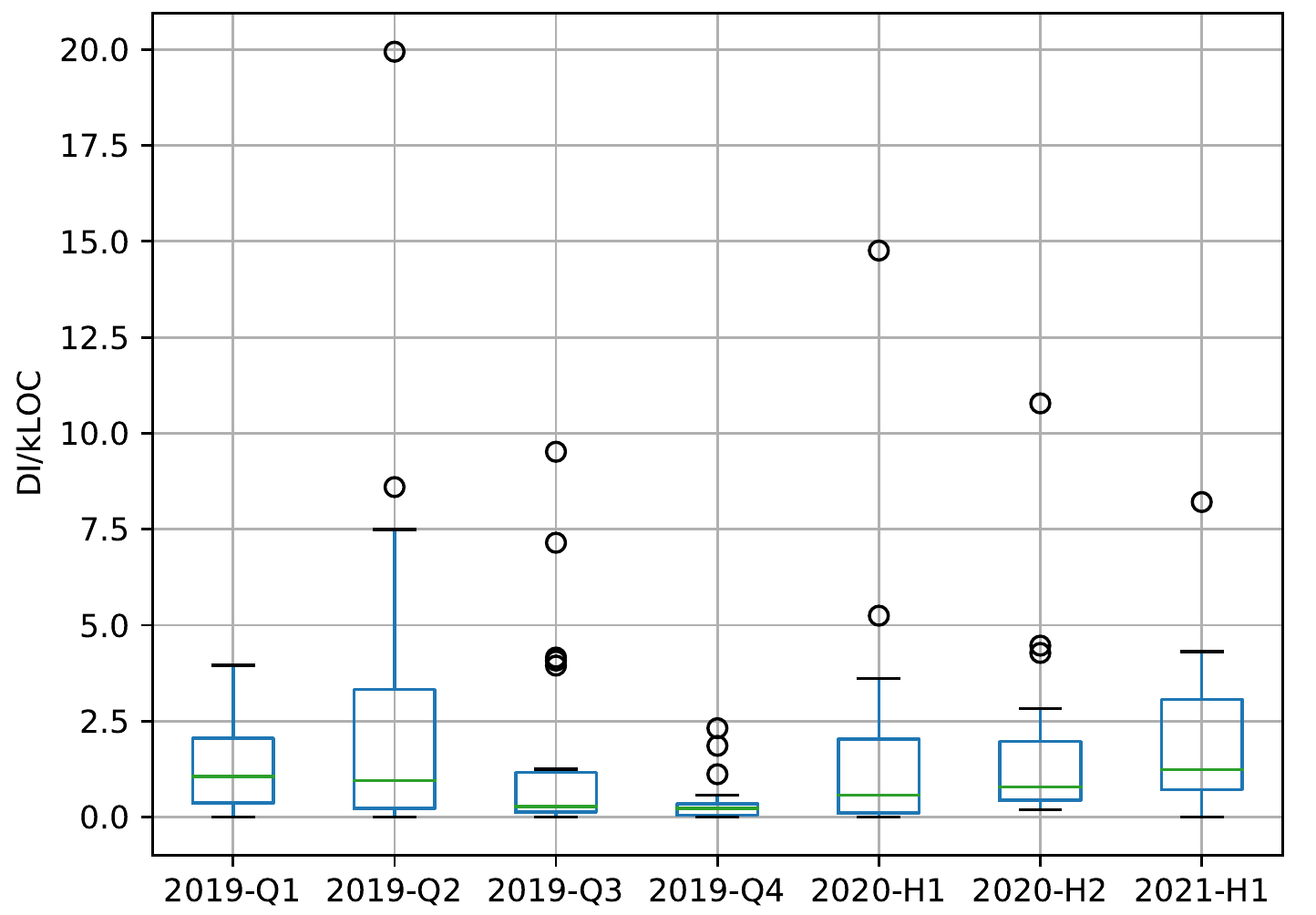}
  \vspace{-1.0ex}
  \caption{Distribution of inspectors' performance}
  \label{FIG:defectFinding}
  \vspace{-2.0ex}
\end{figure}

Where, $N_{maj}$ denotes the number of major defects, $N_{mod}$ for moderate defects, $N_{min}$ for minor defects. Inspectors' outputs are subject to the scale and field of their assignments (samples). 
Figure~\ref{FIG:defectFinding} shows an overview of the inspectors' outputs since 2019. Each retro-inspection involved the same number of inspectors (20). In this boxplot, the green line represents average outputs, the hollow circles indicate `outliers', i.e. inspectors whose outputs (DI/kLOC) are significantly lower or higher than the others in single retro-inspection. We can first observe internal differences from many outliers, then observe external differences from green lines. Inspectors' average DI/kLOC in 2019-Q4 is as low as 0.28, but in 2019-Q2 is as high as 2.01, which can be attributed to one inspector's performance at surprisingly 19.94 DI/kLOC. In a nutshell, inspectors' outputs and engagements vary significantly in retro-inspections.

\paragraph{\textbf{Lesson 3: Organizers' predicament.}}
The organizers have to confront the predicament of coordination in retro-inspections. An organizer ($O_2$) shared his experience, which is typical in retro-inspections. An inspector ($SE_{[1]}$) sent his reports to the inspected division before the deadline but did not receive any response, then he assumed his work was complete. About ten days later, the inspected division contacted the organizer for a number of disagreements, the organizer then forwarded them to the inspector ($SE_{[1]}$), when they both were under a heavy workload. Besides, it is common for inspectors to fail to submit reports on time, which postponed the reporting meeting. Moreover, for the retro-inspection without chief inspectors (arbiters), the organizer, often from management, is not able to deal with technical issues. As a temporary inter-divisional activity, multiple participants do not know each other well in retro-inspections. The organizers are responsible for scheduling meetings only, with no authority on decision-making. As a result, disagreements in retro-inspections may last for a long time, in which all participants' work would be influenced to some extent. 

\vspace{-1.0ex}
\begin{framed}
\vspace{-1.0ex}
\textbf{Lessons:} Despite of its many benefits, retro-inspection also suffers developer's acceptance, inspector's engagement, and organizer's predicament. Therefore, it has not worked completely as expected and needs improvements.
\vspace{-1.0ex}
\end{framed}
\vspace{-1.0ex}

\section{Discussion}
\label{SEC:DIS}
This section discusses the possible dilemmas of retro-inspection and offers some recommendations for its future improvement. 

\subsection{Current dilemmas of retro-inspection}
Unlike testing, code review is more reliant on both code author's and reviewer's skill, experience, and engagement. Peer code reviews are conducted within divisions where code authors and reviewers are acquainted with each other. On the contrary, retro-inspection serves internal audit and has involved multiple roles: (1) management, e.g., executives and organizers, (2) inspectors, (3) stakeholders from the inspected organizations, e.g., project leaders and code authors. Some inspectors indicated that in peer code reviews they paid more attention to critical defects, e.g., design rationality. While in retro-inspections, they tended to focus more on general issues or even coding styles in case of potential disagreements and debates. 
As a temporary inter-divisional practice, the inspected samples and the participants of retro-inspection are sometimes decided in an ad hoc manner, thus multiple roles might not be well-prepared. Besides, retro-inspection may run in rush under the pressure of time limit and outcome evaluation. As a result, retro-inspection might be inefficient sometimes, even became a waste of time. In general, retro-inspection is expected to promote common advance rather than criticism, where the consistency and compromise always outweigh the divergence on experience, preference, and interest among the participants.

\subsection{Recommendations for improvement}
To improve the quality of retro-inspection, we propose the following recommendations. 

\paragraph{\textbf{Recommendation 1: Stricter admission criteria.}}
It is commonly agreed that retro-inspection brings many benefits for various software projects in the enterprise. Although it is required that all the inspected projects should have been reviewed and tested in advance, some did not strictly obey the admission criteria in the past. Moreover, some code files with incomplete functions or missing documents slipped into retro-inspections, which may waste all participants' time and effort. One inspector ($CE_{[1]}$) mentioned that he had received prototyping projects, i.e. the source code of merely interfaces without any implementations, for review. On the contrary, five inspectors ($LE_{[7]}$, $LE_{[1]}$, $SE_{[1]}$, $ME_{[10+]}$, $ME_{[5]}$) indicated that they ever received large scale projects with over 50-80KLOC, which seemed to be a mission impossible for them. It is suggested to carefully choose sample for inspection with stricter admission criteria.

\paragraph{\textbf{Recommendation 2: More adequate preparation.}}
The inspectors' participation in retro-inspections is often decided at the last minute, so are the inspected divisions and samples. Although all the invited inspectors are domain experts and get used to peer code review, they may be new to retro-inspection. There is a need for specific training courses ($ME_{[10+]}$, $ME_{[3]}$) to introduce the inspection process, standard, emphasis and so on, instead of merely a brief kick-off meeting. With a lack of documents ($CE_{[2]}$, $ME_{[10+]}$, $ME_{[3]}$) associated with the inspected project, the inspectors are not able to understand the background ($LE_{[1]}$, $LE_{[1]}$) as well as the samples ($LE_{[4]}$) for inspection. The more information inspectors have about the inspected division and samples, the easier to identify defects and to make suggestions. Hence, once the sample to be inspected is picked, it is important to make adequate preparations for both the inspectors and the inspected division.

\paragraph{\textbf{Recommendation 3: Longer inspection period.}}
Within the limited period of retro-inspection (approx. two weeks), inspector's intensive workload on inspecting samples may be in conflict with his/her routine workload from the division. In such a situation, inspectors always give priority to completing their routine work first. Therefore, the inspection is ostensibly conducted or even incompletely. Taking the second retro-inspection in 2020 as an example, one inspector accepted the invitation, but he did nothing on inspection until a couple of days before the deadline since he was engaged at the time in an urgent task from his division. As a result, he submitted no report, and worse yet, there was no alternative inspector for him. One inspector ($LE_{[1]}$) mentioned that he could inspect only 1/5 of the assignment with 50KLOC. Other inspectors ($LE_{[1]}$, $SE_{[1]}$, $ME_{[5]}$) also indicated that they did not examine all their assignments. Hence, it is recommended to extend the inspection period for another couple of weeks.

\paragraph{\textbf{Recommendation 4: More careful publicity.}}
Publicity is preferred for the inspectors who participated in retro-inspections and the high-performing divisions under inspection. In contrast, the low-performing divisions feel uncomfortable with the inspection results. 
Also the inspectors from the same division may regard each other as competitors in retro-inspections. In this case, peer pressure comes from both within and out of the division ($LE_{[1]}$, $SE_{[2]}$). A number of inspectors ($CE_{[3]}$, $LE_{[1]}$, $LE_{[1]}$, $SE_{[2]}$, $ME_{[5]}$) suggested that it would be better to disclose source code in public instead of projects, divisions, or inspectors. The ultimate objectives of retro-inspection are to promote software quality, strengthen developer's quality awareness, and seek opportunities for inter-division collaborations rather than criticism. Otherwise, the inspection may suffer from a number of negative effects, e.g., disagreements and debates.

\section{Threats to validity}
This section describes the threats to validity of this study as follows.


\paragraph{\textbf{Internal validity.}}
The major threat to internal validity come from the data sources. We are only permitted to access limited data sources due to the enterprise's restrictions. To mitigate this threat, this study involves archive analysis, questionnaire survey, and panel interview for data triangulation. Each type of subject (projects, participants) was carefully selected to ensure they are representative and well-prepared, and the interviewees were pleased to provide insightful comments and recommendations. Another possible threat to internal validity lies in coding when analyzing inspection comments. To this end, we learned the enterprise's specifications, checked the existing labels, inquired developers, and made pilot coding to ensure validity to a large extent. Two authors independently performed coding for each comment, and then three authors collectively checked and fixed all disagreements in consensus meetings. Finally, all the results, findings, and recommendations have been checked and confirmed by the enterprise. 

\paragraph{\textbf{Construct validity.}}
The major threat to construct validity come from measuring inspectors' outputs. The inspected samples, divisions, and inspectors vary across retro-inspections. Also, the code submissions are in the form of modules (more than 2KLOC), thus the assignment for each inspector might be uneven. In order to minimize this threat, we use `DI/kLOC', a common metric in testing, to eliminate the influence of the variety.

\paragraph{\textbf{External validity.}}
As this is a case study reporting on retro-inspection, an enterprise-specific software practice in the industrial context, the experience, lessons, and recommendations may not be applicable out of this enterprise. Other software organizations seeking to conduct retro-inspection or implement other types of QA practices can refer to this enterprise's experience and recommendations, and adapt the practice according to their business and other requirements in the specific context.

\section{Conclusion and future work}
This study reports an empirical investigation of retro-inspection, an altered QA practice, by collecting and analyzing various evidence from a long-period participant observation in one global ICT enterprise. Secured by data triangulation, four major benefits of retro-inspection that outperforms peer code review are identified and empirically confirmed, including identifying complicated and underlying defects, offering more suggestive comments. In addition to recognizing the value of this practice, a few limitations of retro-inspection at this stage are also identified in its preparation and execution according to some participant roles' negative perceptions. We also discuss the recommendations for the improvements of this practice. The experiences on executing retro-inspection in the industrial context can provide the realistic reference value on code quality assurance to other software organizations.

In the future, we are going to continuously observe and report the long-term effects of retro-inspection in more organizations, as well as to research and develop intelligent techniques to advance the support on evaluating and improving code review.

\section*{Acknowledgments}
This work is supported by the National Natural Science Foundation of China (No.62072227), the National Key Research and Development Program of China (No.2019YFE0105500) jointly with the Research Council of Norway (No.309494), the Key Research and Development Program of Jiangsu Province (No.BE2021002-2), and the Intergovernmental Bilateral Innovation Project of Jiangsu Province (No.BZ2020017).

\bibliographystyle{ACM-Reference-Format}
\bibliography{main}

\end{document}